\documentclass[twocolumn,showpacs,preprintnumbers,superscriptaddress,prb]{revtex4-1}

\usepackage{graphicx}
\usepackage{color}
\def\vector#1{\mbox{\boldmath $#1$}}

\usepackage{dcolumn}

\usepackage{bm}

\begin{document}

\preprint{}

\title{Pressure dependence of the magnetic ground states in MnP}

\author{M. Matsuda}
\affiliation{Quantum Condensed Matter Division, Oak Ridge National Laboratory, Oak Ridge, Tennessee 37831, USA}

\author{F. Ye}
\affiliation{Quantum Condensed Matter Division, Oak Ridge National Laboratory, Oak Ridge, Tennessee 37831, USA}

\author{S. E. Dissanayake}
\affiliation{Quantum Condensed Matter Division, Oak Ridge National Laboratory, Oak Ridge, Tennessee 37831, USA}

\author{J.-G. Cheng}
\affiliation{Beijing National Laboratory for Condensed Matter Physics and Institute of Physics, Chinese Academy of Sciences, Beijing 100190, China}

\author{S. Chi}
\affiliation{Quantum Condensed Matter Division, Oak Ridge National Laboratory, Oak Ridge, Tennessee 37831, USA}

\author{J. Ma}
\affiliation{Department of Physics and Astronomy, University of Tennessee, Knoxville, Tennessee 37996, USA}

\author{H. D. Zhou}
\affiliation{Department of Physics and Astronomy, University of Tennessee, Knoxville, Tennessee 37996, USA}

\author{J.-Q. Yan}
\affiliation{Materials Science and Technology Division, Oak Ridge National Laboratory, Oak Ridge, Tennessee 37831, USA}
\affiliation{Department of Materials Science and Engineering, University of Tennessee, Knoxville, Tennessee 37996, USA}

\author{S. Kasamatsu}
\affiliation{Institute for Solid State Physics, University of Tokyo, Kashiwa, Chiba 277-8581, Japan}

\author{O. Sugino}
\affiliation{Institute for Solid State Physics, University of Tokyo, Kashiwa, Chiba 277-8581, Japan}

\author{T. Kato}
\affiliation{Institute for Solid State Physics, University of Tokyo, Kashiwa, Chiba 277-8581, Japan}

\author{K. Matsubayashi}
\affiliation{Institute for Solid State Physics, University of Tokyo, Kashiwa, Chiba 277-8581, Japan}

\author{T. Okada}
\affiliation{Institute for Solid State Physics, University of Tokyo, Kashiwa, Chiba 277-8581, Japan}

\author{Y. Uwatoko}
\affiliation{Institute for Solid State Physics, University of Tokyo, Kashiwa, Chiba 277-8581, Japan}

\date{\today}

\begin{abstract}

MnP, a superconductor under pressure, exhibits a ferromagnetic order below $T\rm_{C}$$\sim$290 K followed by a helical order with the spins lying in the $ab$ plane and the helical rotation propagating along the $c$ axis below $T\rm_{s}$$\sim$50 K at ambient pressure. We performed single crystal neutron diffraction experiments to determine the magnetic ground states under pressure. Both $T\rm_{C}$ and $T\rm_{s}$ are gradually suppressed with increasing pressure and the helical order disappears at $\sim$1.2 GPa. At intermediate pressures of 1.8 and 2.0 GPa, the ferromagnetic order first develops and changes to a conical or two-phase (ferromagnetic and helical) structure with the propagation along the $b$ axis below a characteristic temperature. At 3.8 GPa, a helical magnetic order appears below 208 K, which hosts the spins in the $ac$ plane and the propagation along the $b$ axis. The period of this $b$ axis modulation is shorter than that at 1.8 GPa. Our results indicate that the magnetic phase in the vicinity of the superconducting phase may have a helical magnetic correlation along the $b$ axis.
\end{abstract}


\maketitle

Magnetic fluctuations mediated superconductivity is one of the topical issues in the field of condensed matter physics. In copper oxides, iron pnictides, and heavy fermion superconductors, spin fluctuations that persist in the unconventional superconducting state are generally considered to be closely related to the superconducting pairing mechanism. \cite{dai,fujita,pfleiderer}

The discovery of pressure-induced superconductivity in CrAs, \cite{wu2014,kotegawa} which has the MnP-type structure, showed a new avenue to the search for new superconductors. 
Very recently, bulk superconductivity with transition temperature of $T\rm_{sc}$$\sim$1 K was discovered in MnP by applying a critical pressure ($P\rm_{c}$) of $\sim$7.8 GPa. \cite{cheng}
This is the first observation of superconductivity in the Mn-based compounds.
Since superconductivity emerges in the vicinity of a new, presumably antiferromagnetic phase, it is crucial to elucidate its magnetic structure and clarify the interplay between magnetism and superconductivity in MnP. \cite{norman2}
CrAs and MnP adopt the same orthorhombic structure ($Pnma$). Their difference lies mainly in the dominating magnetic interaction: antiferromagnetic in CrAs and ferromagnetic in MnP. Although both materials show a helical magnetic structure as the ground state at ambient pressure, the spin configuration in CrAs is in proximity to an antiferromagentic phase, while that in MnP is near a ferromagnetic one. The pressure-temperature phase diagram is also different for the two materials. CrAs shows a first order transition to a helical structure with the magnetic modulation vector ($\vector{q\rm_m}$) along the $c$ axis at 260 K, accompanied by a structural expansion, at ambient pressure. With applied pressure, the transition is gradually suppressed and superconductivity appears below $\sim$1.2 K at $\sim$0.35 GPa. The magnetic and superconducting phases coexist between 0.35 and 0.7 GPa. \cite{wu2014,keller,khasanov,shen} The static magnetic order is entirely suppressed at 0.7 GPa and the superconductivity remains. \cite{wu2014,kotegawa,khasanov} CrAs was found to be an unconventional superconductor without a coherence effect in the NQR measurement. \cite{kotegawa2}

\begin{figure}
\includegraphics[width=7.1cm]{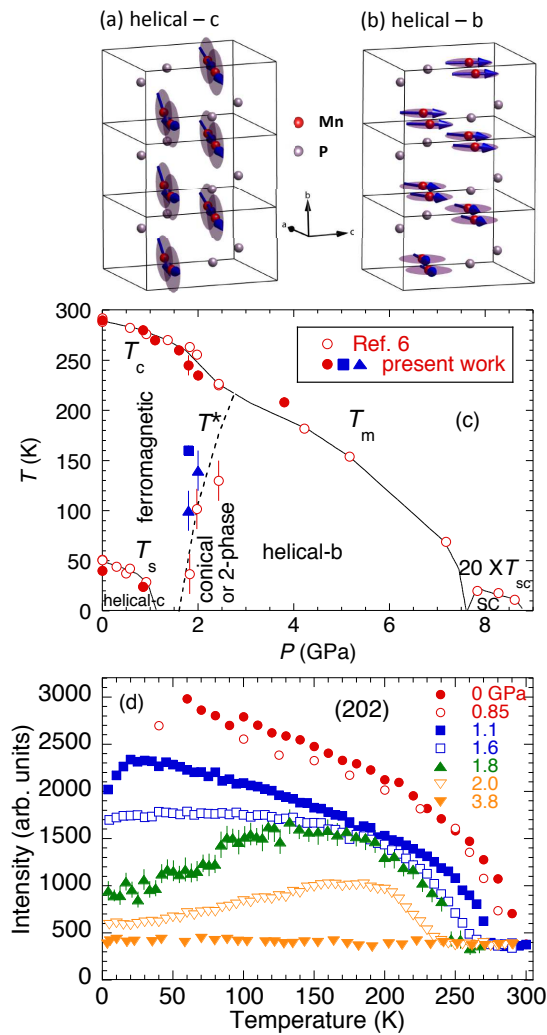}
\caption{(Color online) Helical spin structures under low (helical-$c$) (a) and high pressures (helical-$b$) (b). (c) Temperature-pressure phase diagram. The open and filled symbols represent the data in Ref. \onlinecite{cheng} and the present results, respectively. The filled square is the temperature where the incommensurate magnetic peaks start to develop. The filled triangles are the temperatures where the commensurate magnetic signal starts to decrease. The lines are guides to the eye. (d) Temperature dependence of the (202) Bragg intensity in MnP measured at various pressures. The finite intensity above $T\rm_{C}$ originates from nuclear Bragg intensity, which is used to normalize the data.}
\label{202peak}
\end{figure}
In contrast, MnP shows a ferromagnetic long range order below $T\rm_C$=291 K followed by another transition to a helical structure with $\vector{q\rm_m}\parallel c$ and the easy plane in the $ab$ plane (helical-$c$, Fig. \ref{202peak}(a)) at $T\rm_s$=50 K, \cite{felcher,forsyth,moon} as shown in Fig. \ref{202peak}(c).
With the application of pressure, both magnetic transition temperatures are reduced. At 1.2 GPa, the helical phase is destroyed and the ferromagnetic state becomes the ground state. Above 1.7 GPa, the ferromagnetic phase appears first with decreasing temperature and a new magnetic phase appears below $T^*$, where the ferromagnetic component observed by the ac magnetic susceptibility measurements decreases. Above 3 GPa, no ferromagnetic order was observed down to low temperatures but the resistivity starts to decrease faster below $T\rm_{m}$ and this anomaly in resistivity is suppressed completely around $P\rm_{c}$=8 GPa where superconductivity emerges. The new phase below $T\rm_{m}$ was attributed to an antiferromagnetic state since the magnetic susceptibility is small. Very recently, it was reported from the synchrotron x-ray diffraction measurements that incommensurate peaks are observed at (0,0,1$\pm\delta$) with $\delta\sim$0.25 (in $Pnma$ notation) between 3.17 and 6.43 GPa, which was ascribed to be magnetic in origin. \cite{wang}
However, many characteristics of this intermediate magnetic order remain unknown.

In this Rapid Communication, we report a detailed neutron diffraction study characterizing the pressure evolution of the magnetic structure. In agreement with the previous magnetic susceptibility results, we found that both $T\rm_C$ and $T\rm_s$ gradually decrease with external pressure. The helical order is entirely suppressed at $\sim$1.2 GPa. At 1.8 GPa, the ferromagnetic order develops below $T\rm_C$=250(5) K and then changes to a conical or two-phase (ferromagnetic and helical) state with $\vector{q\rm_m}\parallel b$ below $T^*\sim$150 K. At 3.8 GPa the magnetic ground state has a helical structure without ferromagnetic component.
The results indicate that the superconducting pairing mechanism could be common in CrAs and MnP, although the direction of $\vector{q\rm_m}$ of the helical magnetic state in the vicinity of the superconducting phase is different.

High pressure neutron diffraction measurements were performed on MnP using the triple-axis spectrometer HB-1 and the wide angle diffractometer WAND at the High Flux Isotope Reactor (HFIR) and the time-of-flight diffractometer CORELLI at the Spallation Neutron Source (SNS) at Oak Ridge National Laboratory (ORNL). The single crystals were grown by a modified Bridgman method from a mixture of stoichiometrically equal amounts of Mn and P.
The high pressures below and above 2 GPa were generated with a self-clamped piston-cylinder cell (SCPCC) and a palm cubic anvil cell (PCAC), respectively. The crystal dimensions are 1$\times$1$\times$3 mm$^{3}$ and 2$\times$2$\times$2 mm$^{3}$ for SCPCC and PCAC, respectively.
The SCPCC was made of a Zr-based amorphous alloy. \cite{cell} Fluorinert was chosen as the pressure transmitting medium (PTM). The pressure inside SCPCC was monitored by measuring the lattice constant of a comounted NaCl crystal. We found that the pressure is reduced by a few percent with decreasing temperature from room temperature to 5 K. Therefore, the uncertainty of pressures is considered to be within the symbol size shown in Fig. \ref{202peak}(c).
The PCAC consists of a cluster of six ZrO$_2$ anvils converging onto the center gasket from three orthogonal directions. Using this newly developed PCAC, crystal with several times larger than previously used can be measured under highly hydrostatic condition. \cite{utsumi,abe,cheng14} The sample is immersed in the liquid PTM (fluorinert) contained in a teflon capsule.
The pressure values shown in this paper are determined at room temperature.
The single crystal, which was mounted with $(H0L)$ in the horizontal scattering plane in the pressure cell, was cooled down using a closed cycle refrigerator. Additional experiments in the $(0KL)$ scattering plane were also performed at 3.8 GPa.

\begin{figure}
\includegraphics[width=8.5cm]{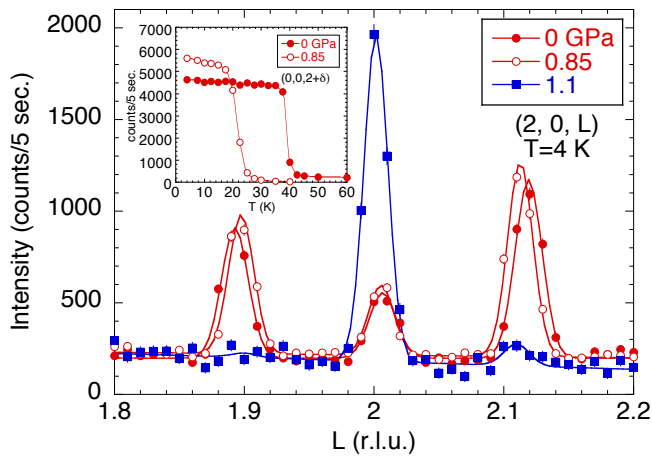}
\caption{(Color online) $L$-scans around the (202) Bragg peak at 4 K as a function of pressure. The solid lines are the results of fits to three Gaussians. r.l.u. stands for reciprocal lattice unit. The insert shows the temperature dependence of the incommensurate magnetic peak (0, 0, 2+$\delta$) at 0 and 0.85 GPa. The solid lines are guide to the eyes.}
\label{spiral}
\end{figure}
Figure \ref{202peak}(d) shows the temperature dependence of (202) Bragg peak intensity at various pressures. 
We confirmed that the (202) Bragg reflection is not affected by the extinction effect so that its nuclear intensity is reliable when compared to the magnetic intensity. 
At ambient pressure, the (202) ferromagnetic intensity develops at $T\rm_C <$290 K. The intensity drops abruptly below $T\rm_s$=40(1) K (not shown). Simultaneously, new magnetic peaks originating from the helical-$c$ structure develop at incommensurate positions that split along the $L$ direction, such as (2, 0, 2$\pm$$\delta$), as shown in Fig. \ref{spiral}. With increasing pressure, $T\rm_C$ and $T\rm_s$ decrease gradually. The incommensurability ($\delta$) decreases slightly at higher pressures. This result suggests that the ferromagnetic interactions become dominant with pressure. At 1.1 GPa, the slight intensity drop of the (202) peak below 20 K and the very small intensity around (2, 0, 2.11) suggest that a small fraction of the sample still remains in the helical phase.

At higher pressures, $T\rm_C$ continues to decrease. The magnetic intensity of the (202) peak also decreases with increasing pressure, indicating that the ferromagnetically ordered Mn moment is further reduced. At 165 K and 2.0 GPa, the magnetic ordered moment is determined to be 0.7(1)$\mu\rm_B$ which is much reduced from the ordered moment of 1.3(2)$\mu\rm_B$ at 60 K at ambient pressure. \cite{supplemental}
The temperature profile of the intensity for the (202) peak also changes above 1.1 GPa.
At 1.6 GPa, the intensity becomes almost constant below 130 K. Such an intensity reduction becomes clearer and takes place at temperature of $\sim$100 K at 1.8 GPa and $\sim$140 K at 2.0 GPa.

\begin{figure}
\includegraphics[width=8.5cm]{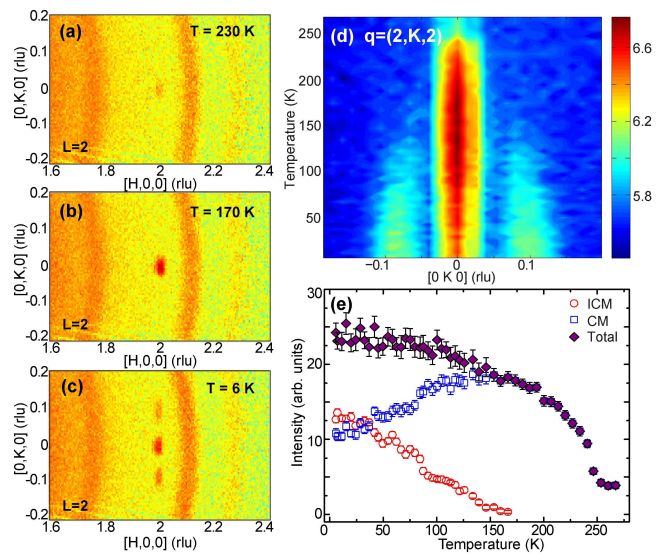}
\caption{(Color online) Intensity contour plot of neutron diffraction at 1.8 GPa in the $(HK2)$ scattering plane measured at 230 (a), 170 (b), and 6 K (c). The arc-shaped signals are powder lines from the Zr-based amorphous pressure cell. (d) Intensity contour plot of the $(2,K,2)$ (-0.2 $\le K\le$ 0.2 r.l.u.) intensities as a function of temperature. (e) Temperature dependence of the (2,0,2) commensurate intensity, sum of the (2,$\pm\delta$,2) incommensurate magnetic intensities, and the total intensity at 1.8 GPa.}
\label{corelli1}
\end{figure}
\begin{figure}
\includegraphics[width=8.5cm]{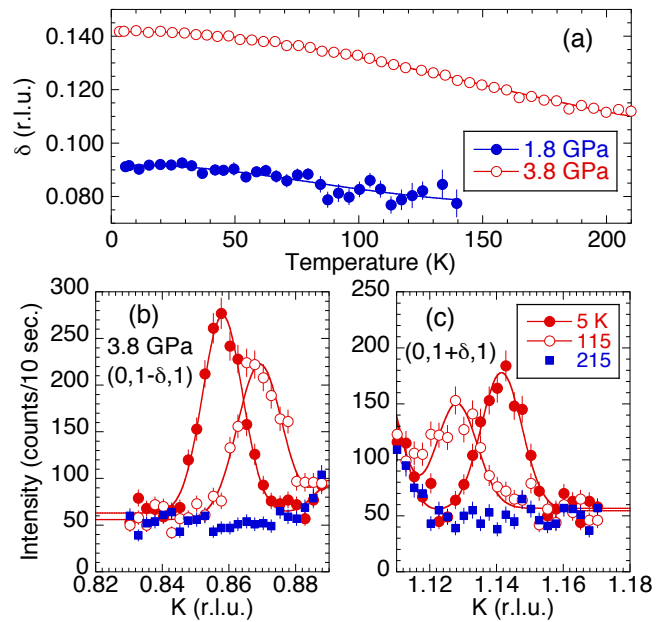}
\caption{(Color online) (a) Temperature dependence of $\delta$ of the incommensurate peaks at (2,$\pm\delta$,2) at 1.8 and 3.8 GPa. (b) and (c) show the incommensurate magnetic peaks at (0,1$-\delta$,1) and (0,1$+\delta$,1), respectively, observed at 3.8 GPa and at 5, 115, and 215 K. The upturn background in (b) and (c) originates from a powder line of the pressure cell. The solid lines are guides to the eye.}
\label{ICM}
\end{figure}

In order to clarify whether additional magnetic orders appear below $T\rm^{*}$, a neutron diffraction experiment was performed at 1.8 GPa using CORELLI.
Figures \ref{corelli1}(a)-(c) show that new incommensurate peaks, which split along $K$ direction, appear below $\sim$150 K. As shown in Fig. \ref{corelli1}(d), the intensities at (2,$\pm\delta$,2) and incommensurability $\delta$ gradually increase with decreasing temperature. 
Figure \ref{corelli1}(e) shows more qualitative results of the (202) and (2,$\pm\delta$,2) peaks. The incommensurate peaks appear below 160(5) K, where the commensurate peak becomes flattened. The total intensity continuously increases and no distinct anomaly was observed. These results strongly indicate that the new peaks are magnetic in origin. $\delta$ increases from 0.080(5) r.l.u. at 125 K to 0.091(1) r.l.u. at 6 K, as shown in Fig. \ref{ICM}(a). The peak widths of both incommensurate and commensurate peaks are sharp and resolution limited, indicating that the magnetic phase below $T^{*}$ is still long-range ordered. A magnetic structure analysis has been performed at 6 K. \cite{supplemental}
The incommensurate magnetic peaks can appear in helical, cycloidal, and spin-density-wave structures. All the possible magnetic structures from the structural symmetry were examined. The observed incommensurate magnetic intensities were found to be best reproduced by an elliptical helical structure which has $\vector{q\rm_m}\parallel b$ and the easy plane in the $ac$ plane (helical-$b$, Fig. \ref{202peak}(b)). We also found that an anisotropic helical model with the $a$ axis moment elongated fits the observed intensities even better.
The fitted magnetic parameters are summarized in Table I. The elliptical helical structure was also reported in the helical-$c$ state at ambient pressure. \cite{forsyth} 
On the other hand, the commensurate magnetic peaks remaining below $T^{*}$ originate from the ferromagnetic component along the $b$ axis, as in the ferromagnetic phase above  $T^{*}$. \cite{supplemental}
From the spin components shown in Table I, the averaged Mn moment at each site is estimated to be 1.5(2) $\mu\rm_B$, which is larger than 1.0(1) $\mu\rm_B$ at the ferromagnetic state. This behavior is very similar to that observed at ambient pressure, where the Mn moment grows from 1.3 $\mu\rm_B$ in the ferromagnetic state to 1.58 $\mu\rm_B$ in the helical-$c$ state. \cite{forsyth}

There are two ways to interpret the magnetic structure below $T^{*}$. One is the single phase model with a conical structure, in which the $b$ component is ferromagnetic and the $ac$ component is helical-$b$. The other is the two-phase model with a ferromagnetic and a helical-$b$ structures. Our neutron diffraction results cannot distinguish the two models. However, the gradual temperature and pressure dependences of $\delta$ and the magnetic intensities suggest that the magnetic state below $T^{*}$ is probably conical and the cone angle becomes larger gradually with decreasing temperature and increasing pressure.
In the conical structure model, the averaged Mn moment and cone angle at 1.8 GPa and 6 K is 1.5(2)$\mu\rm_B$ and 61(3)$^{\circ}$, respectively.
It is noted that the transition temperature $T^{*}$ estimated in this study is larger than that in Ref. \onlinecite{cheng}, as shown in Fig. \ref{202peak}(c). This is probably because even in the conical structure the ferromagnetic component observed in magnetization measurements is still large and does not decrease until the commensurate magnetic peak starts to decrease.
\begin{table}
\caption{$\delta$ and magnetic moments along $b$ ($M_b$), $a$ ($M_a$) and $c$ axes ($M_c$) at 1.8 and 3.8 GPa, where the $ac$ plane spin component has the elliptical helical-$b$ structure. $M_b$ at 1.8 GPa and 6 K was deduced from that at 2.0 GPa and 175 K, where similar (202) intensity is observed.}
\begin{ruledtabular}
\begin{tabular}{cccccc}
$P$ (GPa)&$T$ (K)&$\delta$ (r.l.u.)&$M_b$ ($\mu\rm_B$)&$M_{a}$ ($\mu\rm_B$)&$M_{c}$ ($\mu\rm_B$)\\
\hline
1.8 & 6 & 0.091(1) & 0.7 & 1.5(2) & 1.1(2)\\
3.8 & 5 & 0.141(1) & $-$ & 0.93(10) & 0.76(18)\\
\end{tabular}
\end{ruledtabular}
\end{table}

At 3.8 GPa, the intensity of the (202) reflection does not change with decreasing temperature down to 3 K, as shown in Fig. \ref{202peak}(d). This indicates that the ferromagnetic order is completely suppressed. However, the neutron diffraction experiment in the $(0KL)$ scattering plane on HB-1 clearly shows incommensurate magnetic peaks at (0,1$\pm\delta$,1) below 208(5) K [Figs. \ref{ICM}(b) and (c)]. The intensity at (0,1-$\delta$,1) is larger than that at (0,1+$\delta$,1) because of the magnetic form factor and ellipticity of the helical structure. $\delta$ is found to be 0.141(1) r.l.u. at 5 K, which is larger than the value at 1.8 GPa, and it becomes reduced with increasing temperature, as shown in Fig. \ref{ICM}(a). The magnetic structure analysis indicates that the magnetic structure is helical-$b$. \cite{supplemental} As shown in Table I, the averaged magnetic moment is estimated to be 0.84(14)$\mu\rm_B$ at 5 K. Therefore, the conical or two-phase structure below $T^{*}$ is considered to change continuously to helical-$b$ state below $T\rm_m$.
The incommensurability $\delta$(=0.141) at 3.8 GPa is similar to that in CrAs ($\delta\sim$0.14) at 0.88 GPa, \cite{shen} which is located in the border of the bulk superconducting phase.
Furthermore, Shen $et$ $al.$ reported that the easy plane changes from $ab$ to $ac$ in the vicinity of the superconducting phase. \cite{shen} This behavior is also very similar to that in MnP. It is noted that in our measurements no signals were observed at (0,0,1$\pm\delta$) with $\delta\sim$0.25 reported in Ref. \onlinecite{wang}, where both the direction and the value are different from the present results, suggesting that the signals at (0,0,1$\pm\delta$) may be sample dependent or come from the surface.

What is the origin of the new magnetic phase below $T\rm^{*}$ and $T\rm_{m}$?
According to previous inelastic neutron scattering measurements in MnP at ambient pressure, \cite{tajima,todate,yano} spin-wave excitations extend at least up to 75 meV. The magnetic interactions up to sixth nearest-neighbor are relevant. $J_1$ (the nearest-neighbor interaction), $J_2$ (the second-neighbor interaction), $J_5$ (the fifth-neighbor interaction) and $J_6$ (the sixth-neighbor interaction) are mostly dominant with $J_2$ being antiferromagnetic and the others ferromagnetic. \cite{yano}
The successive magnetic transitions at ambient and low pressures indicate competing ferromagnetic and antiferromagnetic interactions. With applying pressure, balance of these interactions can be changed, leading to evolution of magnetic ground states from the helical-$c$ to ferromagnetic to the helical-$b$ spin configurations.
It is important to clarify the mechanism of the change of the direction of $\vector{q\rm_m}$ from the $c$ to $b$ axis, as well as the change of the magnetic anisotropy from the $bc$ to $ac$ plane.

At ambient pressure, MnP is considered to be a localized $d$ electron spin system of Mn, interacting with itinerant $s$ electrons of P, which is reproduced by the $s$-$d$ model. Applying pressure causes more enhanced orbital overlap between Mn atoms, which gives rise to more itinerancy. The density-functional theory (DFT) calculation shows that the pressure gradually reduces the Mn moments and finally leads to a nonmagnetic state. \cite{gercsi}
The Mn moment estimated at 3.8 GPa and 5 K is 0.84(14)$\mu\rm_B$, which suggests that the lattice compression is very close to the critical regime where the large $d$-$d$ overlapping makes the system more itinerant and the spontaneous magnetization does not occur. \cite{gercsi}
In Ref. \onlinecite{gercsi}, only collinear spin structures (one ferromagnetic and three antiferromagnetic structures) are considered as potential ground states. The ferromagnetic structure was found to be most stable in MnP until the magnetic moment disappears with compression. This result does not exclude the possibility of a noncollinear magnetic state between ferromagnetic and itinerant nonmagnetic states. Further theoretical studies on the magnetic state in the vicinity of the itinerant nonmagnetic state are desirable.

In summary, the pressure-temperature magnetic phase diagram in MnP is constructed based on the neutron diffraction study under pressure.
The magnetic ground state changes from being helical-$c$ at ambient pressure to ferromagnetic at $\sim$1.2 GPa. Above 1.5 GPa, the conical or two-phase structure with $\vector{q\rm_m}\parallel b$ appears and gradually changes to the helical-$b$ structure, which probably retains up to the vicinity of the superconducting phase. The helical magnetic structure with the $ac$ easy plane may be common in the vicinity of the superconducting phase in MnP and CrAs, although the direction of $\vector{q\rm_m}$ is different and the nearest-neighbor coupling is close to ferromagnetic and antiferromagnetic in MnP and CrAs, respectively.
Thus, MnP may provide an opportunity to study the mechanism of unconventional superconductivity in the vicinity of the helical magnetic phase close to the ferromagnetic state.

\begin{acknowledgments}
We are grateful to Drs. Kazuki Komatsu (Univ. of Tokyo) and Yoshihiko Yokoyama (IMR, Tohoku Univ.) for use of the pressure cell. Research conducted at ORNL's HFIR and SNS was sponsored by the Scientific User Facilities Division, Office of Basic Energy Sciences, US Department of Energy. This study was supported in part by the U.S.-Japan Cooperative Program on Neutron Scattering. JGC is supported by the MOST and NSF of China (Grants Nos. 2014CB921500, 11574377, and 11304371), and the Strategic Priority Research Program (B) of the Chinese Academy of Sciences (Grants No. XDB07020100). HDZ acknowledges the support from NSF-DMR-1350002. JQY acknowledges the support by the U.S. Department of Energy, Office of Science, Basic Energy Sciences, Materials Sciences and Engineering Division. YU was supported by JSPS KAKENHI (Grant No.15H03681).
This manuscript has been authored by UT-Battelle, LLC under Contract No. DE-AC05-00OR22725 with the U.S. Department of Energy. The United States Government retains and the publisher, by accepting the article for publication, acknowledges that the United States Government retains a non-exclusive, paid-up, irrevocable, world-wide license to publish or reproduce the published form of this manuscript, or allow others to do so, for United States Government purposes. The Department of Energy will provide public access to these results of federally sponsored research in accordance with the DOE Public Access Plan(http://energy.gov/downloads/doe-public-access-plan).
\end{acknowledgments}

\end{document}